\pdfoutput=1

\documentclass[letterpaper, 10 pt, conference]{ieeeconf}  

\IEEEoverridecommandlockouts                              
\usepackage[utf8]{inputenc}
\usepackage[T1]{fontenc}
\usepackage{graphicx}
\usepackage{cite}
\usepackage[colorlinks  = true,
            citecolor   = blue,
            linkcolor   = blue,    
            urlcolor    = blue,   
            breaklinks  = true,    
            pdfauthor   = {Will\ Coon},
            pdftitle    = {},
            pdfsubject  = { },
            pdfkeywords = { },
            plainpages  = false,
           ]{hyperref}
\usepackage{dsfont}

\newcommand{\figu}{Fig.}

\title{\LARGE \bf
Getting More from Less: Transfer Learning Improves Sleep Stage Decoding Accuracy in Peripheral Wearable Devices
}

\author{William G. Coon$^{1,}$$^{2}$, Diego Luna$^{1}$, Akshita Panagrahi$^{1}$, Matthew Reid$^{3}$, Mattson Ogg$^{1}$
\thanks{*This work was supported by the Johns Hopkins University Applied Physics Laboratory.}
\thanks{$^{1}$Johns Hopkins University Applied Physics Laboratory,
        11100 Johns Hopkins Rd., Laurel, MD 20723, USA
        {\tt\small william.coon at jhuapl.edu}}%
\thanks{$^{2}$Johns Hopkins University Whiting School of Engineering, 11100 Johns Hopkins Rd., Laurel, MD 20723, USA}
}

\begin{document}

\maketitle
\thispagestyle{empty}
\pagestyle{empty}

\begin{abstract}
Transfer learning, a technique commonly used in generative artificial intelligence, allows neural network models to bring prior knowledge to bear when learning a new task. This study demonstrates that transfer learning significantly enhances the accuracy of sleep-stage decoding from peripheral wearable devices by leveraging neural network models pretrained on electroencephalographic (EEG) signals. Consumer wearable technologies typically rely on peripheral physiological signals such as pulse plethysmography (PPG) and respiratory data, which, while convenient, lack the fidelity of clinical electroencephalography (EEG) for detailed sleep-stage classification. We pretrained a transformer-based neural network on a large, publicly available EEG dataset and subsequently fine-tuned this model on noisier peripheral signals. Our transfer learning approach improved overall classification accuracy from 67.6\% (baseline model trained solely on peripheral signals) to 76.6\%. Notable accuracy improvements were observed across sleep stages, particularly lighter sleep stages such as REM and N1. These results highlight transfer learning's potential to substantially enhance the accuracy and utility of consumer wearable devices without altering existing hardware. Future integration of self-supervised learning methods may further boost performance, facilitating more precise, longitudinal sleep monitoring for personalized health applications.

\end{abstract}

%
\section{INTRODUCTION}
\label{intro}

\begin{figure*}[thpb]
      \centering
      \includegraphics[width=1\textwidth]{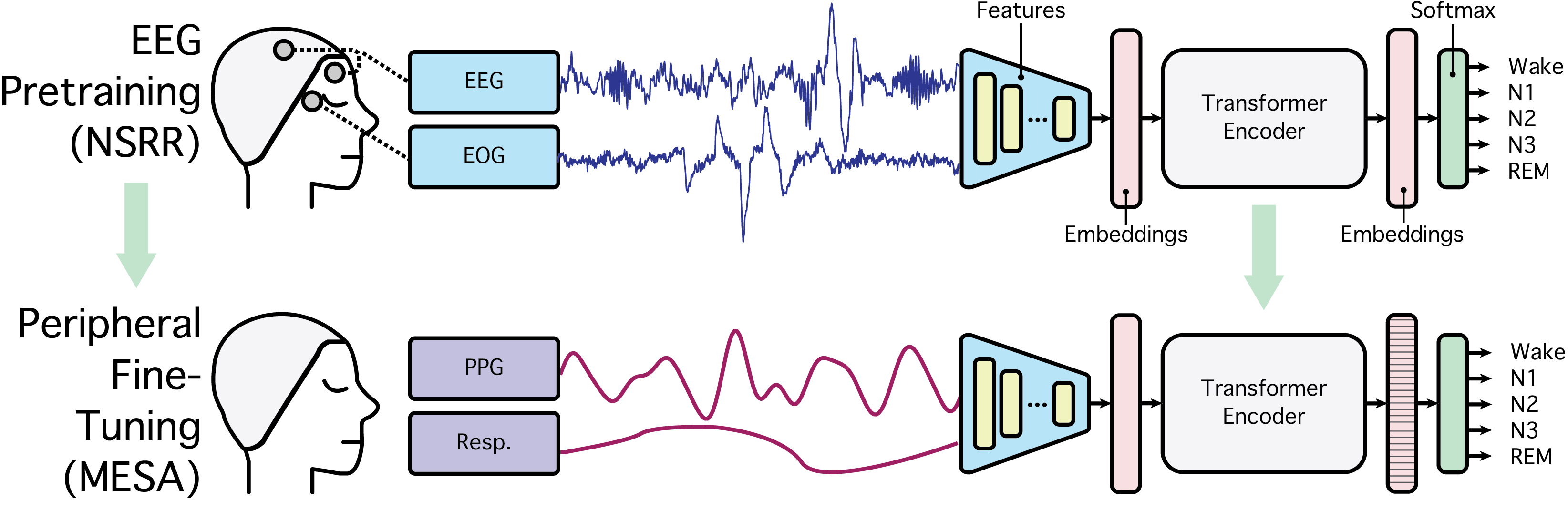}
      \caption{\textbf{Transfer Learning approach.} Transformer models (see Fig. \ref{arch} for full details of model architecture) are first pretrained on $10,897$ recordings of EEG and EOG extracted from full overnight polysomnography (PSG) in a total of $9,013$ individuals. This encourages the model to learn generalizable representations of sleep structure that optimally enable accurate classification of sleep in unseen EEG+EOG data. In the second (fine-tuning) stage, the pretrained model is further trained to predict sequences of sleep stages, but from non-EEG/EOG signals instead (pulse / heart rate / heart rate variability from pulse pleysmography (PPG), and respiration from thoracic piezoelectric belts, both included in full PSG). The final model is more accurate at sleep stage classification using these wearables-accessible signals than a model simply trained from scratch on these signals.}
      \label{archcxr}
    \end{figure*}

Consumer wearable devices such as Oura rings \cite{ouraring} and Apple Watch \cite{applewatch} increasingly provide sleep stage hypnograms, which visualize sleep dynamics across the night. Typically, these devices decode sleep stages from peripheral physiological data like heart rate, heart rate variability, pulse plethysmography, and accelerometry. However, it is widely acknowledged that such peripheral signals lack sufficient fidelity to support reliable classification into the clinically recognized five-stage taxonomy (Wake, N1, N2, N3, and REM). Consequently, many commercial devices simplify the classification into fewer categories (e.g., Wake, NREM, REM), limiting their utility for research and clinical applications (but see \cite{korkalainen2020deep} for some excellent work in this area). This limitation has constrained researchers' ability to leverage the vast potential for large-scale data mining and longitudinal monitoring offered by the widespread adoption of consumer wearable technology. Thus, developing methods to extract higher-fidelity sleep stage information from wearable sensors would offer significant scientific and clinical benefits.

Concurrent with the rapid adoption of wearable sleep monitoring devices, recent advances in artificial intelligence have significantly accelerated the sophistication of sleep analytics. Techniques such as self-supervised learning (SSL), particularly when combined with pretrained transformer-based neural networks, have enabled researchers to leverage large, publicly available datasets to develop automated sleep-stage decoding models \cite{thapa_sleepfm_2024,ogg_self-supervised_2024,coon_laying_2024}. These pretrained "foundation models" acquire generalized representations of sleep structure during pretraining without the need for manually annotated data. Consequently, they require only limited study-specific data during fine-tuning to achieve high classification accuracy tailored to individual experimental needs \cite{ogg_self-supervised_2024,thapa_sleepfm_2024}.

SSL shares conceptual similarities with transfer learning, an established approach wherein a model pretrained on one dataset is fine-tuned on another. The primary distinction lies in SSL's ability to learn directly from unlabeled data by, e.g., predicting randomly masked data segments based on surrounding context or leveraging contrastive learning techniques, whereas traditional transfer learning requires labeled datasets for supervised pretraining. Both methods have already demonstrated substantial utility, particularly when the pretrained and fine-tuned data modalities closely match—for example, training and testing on audio speech or language data \cite{hsu_hubert_2021, devlin_bert_2019}, or training on EEG sleep recordings \cite{guillot2021robustsleepnet}, with or without subsequent fine-tuning on signals obtained from EEG devices with different sensor configurations \cite{coon2024laying}, such as forehead-mounted EEG patches.

Here, we hypothesize that sleep stage classifiers pretrained on high-fidelity EEG recordings may provide substantial performance improvements when fine-tuned on peripheral physiological signals from wearable devices. We reason that a model trained on EEG data will internalize robust representations of sleep architecture, facilitating enhanced interpretation of less direct and inherently noisier peripheral signals. To test this hypothesis, we utilized data from the National Sleep Research Resource \cite{zhang_national_2018} and present evidence indicating that pretrained transformer-based models can effectively transfer high-fidelity EEG-derived knowledge to improve sleep stage decoding from peripheral wearables. Thus, our approach demonstrates the feasibility of using modern artificial intelligence to derive more detailed and accurate sleep hypnograms from peripheral sensor data, effectively enabling researchers and clinicians to “get more from less.”

\section{METHODS}
\label{meth}

\subsection{Pretraining Data}

 We assembled a dataset of 10,897 sleep sessions from 9,013 individuals from public sources hosted on the National Sleep Research Resource \cite{zhang_national_2018} (NSRR) for pre-training. Details of the specific NSRR datasets used are available in Ogg and Coon \cite{coon2024laying}. From each sleep session, we extracted a central EEG channel (i.e., C3 or C4 from the International 10-20 System), resampled to a 100Hz sampling rate and normalized by subtracting the median and scaling to achieve an interquartile range (IQR) of 1.0, truncated to fall within ±20 IQR. Preprocessing steps were conducted using MNE-Python \cite{gramfort_meg_2013}. Processed signals were then segmented into 30-second epochs aligned with corresponding sleep-stage annotations (according to standardized scoring guidelines (e.g., AASM; \cite{silber_visual_2007}), retaining only epochs assigned a sleep stage label. Training examples were constructed by concatenating sequences of 101 consecutive epochs, advancing in increments of 25 epochs.

\subsection{Training and Validation Data}
 For transfer learning model training and validation, sleep sessions from $1,559$ subjects were extracted from the Multi-Ethnic Study for Artherosclerosis \cite{chen_racialethnic_2015} hosted by the NSRR \cite{zhang_national_2018}.  Ages ranged from 54 to 90 years old.  Subjects’ self-reported racial/ethnic background resulted in a distribution of: 36\% White, 28\% Black/African American, 24\% Hispanic and 12\% Asian. Fifty-three percent (53\%) of these subjects identified as female. For each sleep session, we extracted time series data from two peripheral (non-CNS) channels: Abdomen (Sensor: Compumedics Inductive Respiratory Band) and Pulse Plethysmography (Sensor: Nonin 8000). Signals were resampled and normalized using the same method as the EEG pretraining data. Training examples were also generated in the same way (101-epoch sequences of 30-s sleep signals, stepped in 25-epoch strides for approximately 4x oversampling). 
 
 Fine-tuning data were split into an approximately 90\% ($1,398$ subjects) corpus for training with the remaining 10\% ($161$ subjects) reserved as a final unseen, held-out test set for external validation. To monitor model performance during fine-tuning, the train set was further split into a second 90/10 training/\textit{internal}-validation split (i.e., $1,398$ subjects' data were submitted to the model for training, with 90\% of \textit{those} data used as a training set and the remaining 10\% as a validation \textit{during} training; the additional $161$ held out for the \textit{first} 90/10 split were used as an unseen, external validation set that neither the pretraining nor transfer learning models ``saw'' prior to evaluation).


\begin{figure}[thpb]
      \centering
      \includegraphics[width=.7\columnwidth]{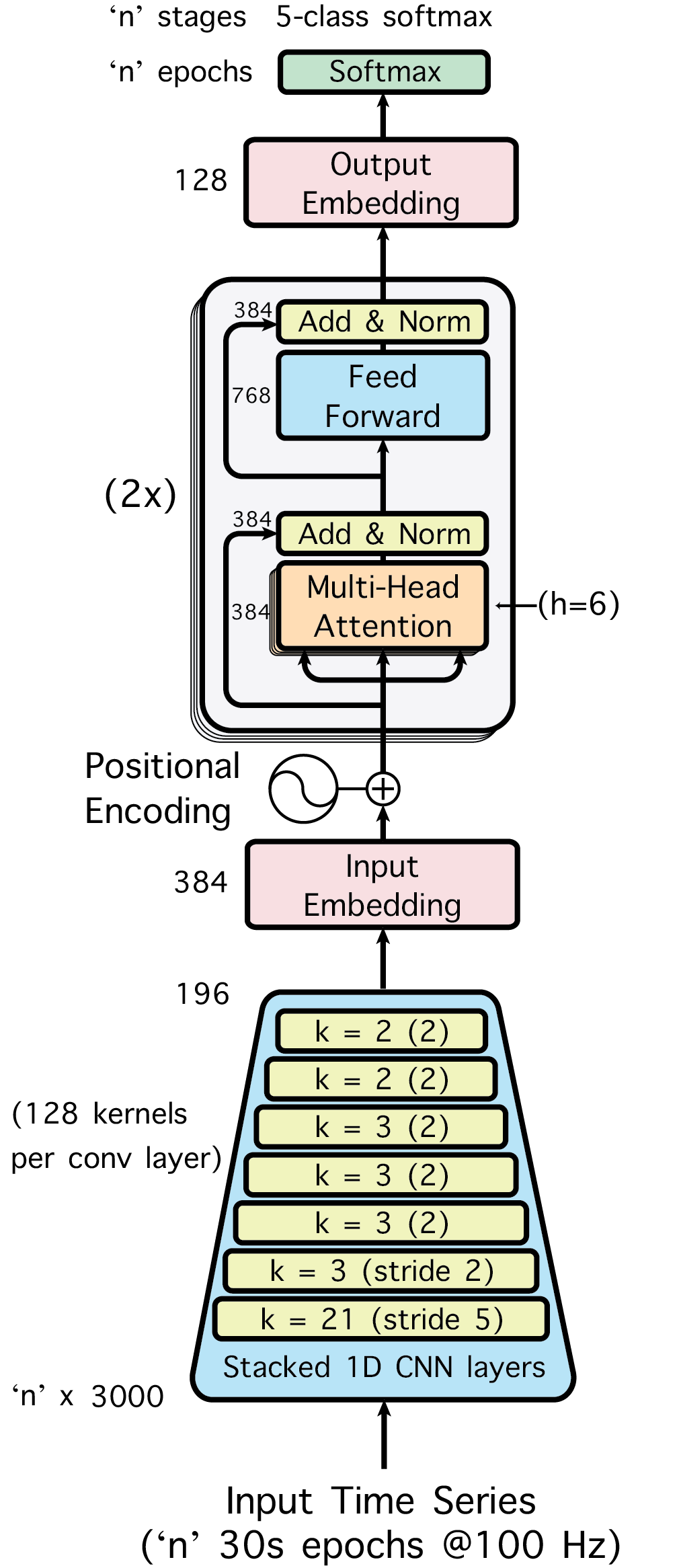}
      \caption{\textbf{Architecture of the transformer-based sleep staging model.} The model is trained to map input time series data to a sequence of sleep stages. A stack of 1D CNNs first extracts features from the raw time series, and is followed by a linear projection to an input embedding, positional encoding (as in Vaswani et al., 2017 (\cite{vaswani2017attention})), multiple transformer encoder layers, another projection to an output embedding, and a final softmax layer for classifying sleep-stage labels (Wake, N1, N2, N3, REM) for each 30-second epoch in the input time series.  The parameters needed to exactly reproduce this in PyTorch are included in the figure, here showing the configuration of the small model variant.}
      \label{arch}
    \end{figure}

\subsection{Model Architecture \& Training}
\label{algo}
    
The model architecture (illustrated in \figu~\ref{arch}) began with a series of seven sequential one-dimensional convolutional layers, to which input time series were submitted for feature extraction and subsequent processing in later network layers. Each convolutional layer featured 128 output channels, with kernels sized [21, 3, 3, 3, 3, 2, 2] and stride lengths of [5, 2, 2, 2, 2, 2, 2]. Following each convolutional step, we applied layer normalization and employed GELU activation functions. Subsequently, the convolutional output was projected through a linear transformation, expanding the dimensionality from 128 to 512, followed again by GELU activation. Output from this layer was then supplemented with positional encoding information.

The position-encoded representations were input into a stack of four transformer encoder layers. Each transformer layer incorporated eight attention heads, a feed-forward intermediate dimension of 768 units, GELU activations, and dropout regularization at a probability of 0.05. After transformer processing, adaptive average pooling in one dimension temporally aligned the transformer outputs with the input epoch labels (consisting of 101 epochs per sequence). The pooled outputs underwent a further linear transformation to yield embeddings of 128 dimensions, activated again using GELU, before passing through the final output layer. 

Overall, the complete model comprised around 3.9 million trainable parameters, with a storage footprint of approximately 43.2 MB. Training spanned 50 epochs, each epoch representing a complete traversal through the entire pretraining dataset. We used batches containing 16 sequences each, optimizing via the Adam algorithm with a linear learning rate schedule as follows: learning rate was initially set at 0.00001 and linearly increased to 0.000375 over the initial 10 epochs.Through error backpropagation, the categorical cross-entropy ($H(P)$) between labeled classes was minimized as:

\begin{equation}
{H(P)} = -\frac{1}{N}\sum_{i=1}^{N}\sum_{c=1}^{C}\mathds{1}_{y_i \in C_c} \log p [y_i \in C_c],
\end{equation}
\\   
\noindent with entropy $P$, number of examples $N$, number of classes $C$ (here, $C=5$ sleep-stage/wake labels), and $y_{i}$ a single example (30-second epoch sequence).  All performance data reported in this manuscript were derived from the (unseen) validation data. Model selection was based on the minimum validation loss achieved throughout training.

\subsection{Transfer Learning}

Transfer learning proceeded in a 2-step process (\figu~\ref{archcxr}). In the first step, the transformer model was initialized with noise weights and trained on EEG/EOG data from the training corpus (see Coon \& Ogg 2024 \cite{coon_laying_2024} for full details on EEG sleep stage classifier model training, and performance results using EEG alone).  In the second step, that pretrained model is \textit{further} trained for 40 more epochs (peak learning rate 0.000025 after 15 epochs; all weights allowed to update- no frozen layers), but using PPG and respiration signals as inputs instead of EEG/EOG. This allows the model to transfer its representation of sleep structure to the task of learning to predict sleep stages from physiologically noisier, peripheral nervous system signals (pulse, breathing). A baseline model was also trained without pretraining (i.e., initialized with noise and trained solely on PPG/respiration signals to predict sleep stages).  This allowed quantification of any performance boost conferred by the pretraining step.

\section{RESULTS}
\label{results}

\begin{figure}[thpb]
      \centering
      \includegraphics[width=0.95\columnwidth]{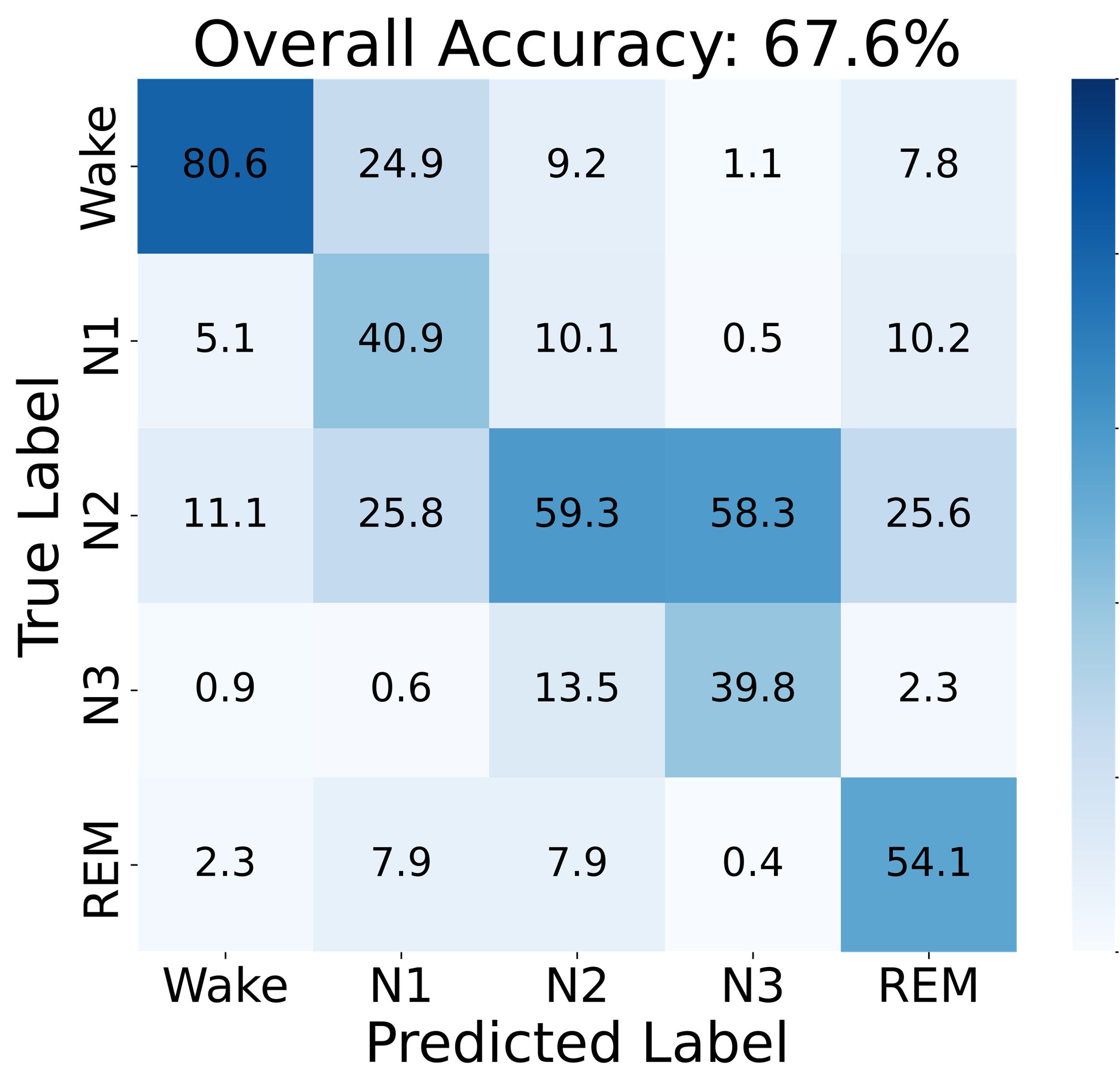}
      \label{woxfer}
   \end{figure}

\begin{figure}[thpb]
      \centering
      \includegraphics[width=0.95\columnwidth]{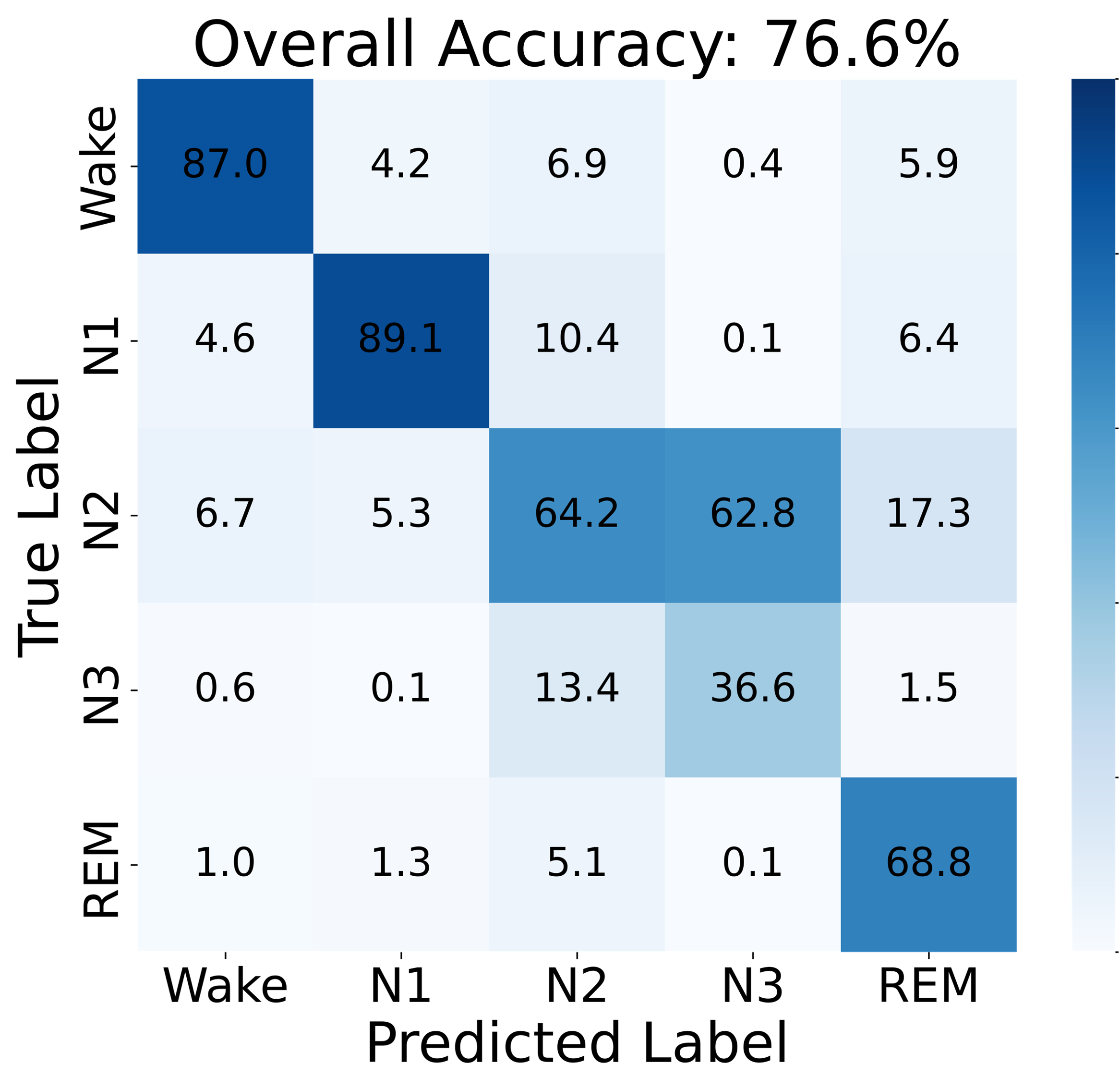}
      \caption{\textbf{Model Performance with and without Transfer Learning}. \textit{TOP:} Confusion matrix showing final performance of the baseline model, which was trained from scratch to predict sleep stage from peripheral physiological signals (pulse, respiration), on the external validation set. Performance is substantially lower than the observed (approximately) 82\% agreement between and within expert sleep scorers visually interpreting full PSG signals \cite{danker-hopfe_interrater_2009}. \textit{BOTTOM:} Model performance when pretrained on over ten thousand recordings using EEG and EOG signals as input, and then transfered to the peripheral signal decoding task via fine-tuning. Performance is substantially better than the baseline model. }
      \label{wxfer}
   \end{figure}

Model performance (as assessed by validation accuracy) reached a loss/accuracy asymptote after approximately 40 passes through the training data, in all cases, and the best-performing model (of the 40 possible model weight sets produced by 40 epochs of training) was chosen for evaluation each time.  Final accuracies  varied substantially.  The baseline model, trained from scratch solely on pulse PPG and respiration data, attained a peak accuracy of 67.6\% overall (\figu~\ref{woxfer}). Its highest accuracy was for Wake (80.6\%), lowest for N3 (39.8\%), and achieved 40.9\% for N1, 59.3\% for N2, and 54.1\% for REM.  In contrast, the transfer learning model (\figu~\ref{wxfer}) achieved a final overall accuracy of 76.6\% (Wake: 87.0\%, N1: 89.1\%, N2: 64.2\%, N3: 36.6\%, and REM: 68.8\%).  

\section{DISCUSSION}
\label{disc}

In this work, we demonstrate compelling evidence that pretraining a deep neural network on large-scale, high-fidelity physiological signals can substantially enhance its ability to decode the same underlying phenomena from lower-fidelity, noisier sources. Specifically, we pretrained a transformer-based neural network on EEG, which provides a high-resolution depiction of human sleep architecture, and then fine-tuned it on peripheral physiological data—pulse plethysmography (PPG) and respiratory signals—that are more easily and commonly measured but contain less direct information about central nervous system sleep dynamics.

Our results indicate a notable performance boost conferred by this transfer learning approach. When compared to a baseline model trained solely on peripheral signals, the pretrained model showed an improvement in overall accuracy from 67.6\% to 76.6\%. Notably, accuracy improved substantially across multiple sleep stages, with particular gains seen in lighter sleep stages (REM and N1). This finding underscores the value of the neural representations of sleep structure internalized by the model during EEG-based pretraining. Such representations appear to significantly inform and enhance the interpretation of the noisier, peripheral signals encountered during subsequent fine-tuning.

Importantly, both training phases in our study employed supervised learning, necessitating labeled data for model development. However, the substantial demands imposed by data labeling could potentially be mitigated through self-supervised learning (SSL) methods. SSL pretraining techniques, which do not require annotated datasets, can harness vast quantities of unlabeled data, thus alleviating the primary bottleneck posed by human annotation constraints. Leveraging SSL may enable even greater improvements in decoding accuracy, opening pathways to utilize larger datasets and potentially further enhancing the precision of sleep-stage classification from peripheral wearables.

Wearable devices hold exceptional promise for naturalistic, long-term monitoring of health states, crucial for personalized health baselining and precision medicine applications. However, despite their proliferation and convenience, current consumer sleep wearables typically rely on peripheral physiological signals and thus exhibit significantly lower accuracy compared to EEG-based clinical monitoring systems. Moreover, inconsistencies in reported results among commercially available wearables create confusion and diminish trust among users, ultimately limiting the utility of these potentially powerful technologies.

The transfer learning approach outlined here presents a practical solution to this challenge. By enhancing the accuracy of peripheral devices without modifying the hardware itself, such methods allow existing wearable devices, characterized by long battery life and user comfort, to achieve EEG-like precision in sleep-stage classification. Future research leveraging self-supervised learning for pretraining could further elevate the performance of these systems, truly enabling us to "get more from less," thereby maximizing the benefits of widely adopted consumer wearable technologies.

\addtolength{\textheight}{-12cm}   







\bibliographystyle{IEEEtran}
\bibliography{biblio}

\end{document}